\documentclass[authornum,preprint,review,12pt]{elsarticle}

\usepackage{lineno,hyperref}
\modulolinenumbers[5]

\usepackage{amsmath,amssymb,amsfonts}
\usepackage{algorithmic}
\usepackage{graphicx}
\usepackage{textcomp}
\usepackage{multirow}
\usepackage{bm}
\usepackage{bbm}
\usepackage{subfigure}
\usepackage{xcolor}
\usepackage{pdflscape}
\journal{Journal of \LaTeX\ Templates}

\begin{document}

\begin{frontmatter}

\title{A Decomposition-Based Hybrid Ensemble CNN Framework for Driver Fatigue Recognition}

\author[address1]{Ruilin Li}
\ead{ruilin001@e.ntu.edu.sg}

\author[address2]{Ruobin Gao}
\ead{gaor0009@e.ntu.edu.sg}

\author[label3,address1]{P. N. Suganthan\corref{mycorrespondingauthor}}
\cortext[mycorrespondingauthor]{Corresponding author}
\ead{p.n.suganthan@qu.edu.qa}

\address[address1]{School of Electrical and Electronic Engineering, Nanyang Technological University, Singapore}
\address[address2]{School of Civil and Environmental Engineering, Nanyang Technological University, Singapore}
\address[label3]{KINDI Center for Computing Research,
College of Engineering, 
Qatar University,
Doha, Qatar}

\begin{abstract}
Electroencephalogram (EEG) has become increasingly popular in driver fatigue monitoring systems. Several decomposition methods have been attempted to analyze the EEG signals that are complex, nonlinear and non-stationary and improve the EEG decoding performance in different applications. However, it remains challenging to extract more distinguishable features from different decomposed components for driver fatigue recognition. In this work, we propose a novel decomposition-based hybrid ensemble convolutional neural network (CNN) framework to enhance the capability of decoding EEG signals. Four decomposition methods are employed to disassemble the EEG signals into components of different complexity. Instead of handcraft features, the CNNs in this framework directly learn from the decomposed components.  
In addition, a component-specific batch normalization layer is employed to reduce subject variability. Moreover, we employ two ensemble modes to integrate the outputs of all CNNs, comprehensively exploiting the diverse information of the decomposed components. Against the challenging cross-subject driver fatigue recognition task, the models under the framework all showed superior performance to the strong baselines. Specifically, the performance of different decomposition methods and ensemble modes was further compared. The results indicated that discrete wavelet transform-based ensemble CNN achieved the highest average classification accuracy of 83.48\% among the compared methods. The proposed framework can be extended to any CNN architecture and be applied to any EEG-related tasks, opening the possibility of extracting more beneficial features from complex EEG data.
\end{abstract}


\end{frontmatter}


\section{Introduction}
\label{sec:introduction}
\textcolor{black}{Driver fatigue is one of the most common and serious hazards on the road and thus a threat to the safety of human life in the world. It often produces drowsiness or even causes short sleep episodes, making drivers less able to recognize potential hazards. Based on the statistics of the causes of traffic accidents in various countries, fatigue driving occupies a large proportion \cite{national2011traffic}. Hence, effectively and efficiently monitoring the driver's fatigue to avoid transportation accidents has become one of the most concerning research topics for automotive safety engineering. In recent years, many researchers have attempted to exploit physiological signals to perform fatigue detection, including but not limited to eye movement signals \cite{HE2017473}, heart rate \cite{8520803}, facial expression \cite{9660782} and \textit{etc.} Among different physiological signals, electroencephalogram (EEG) that can directly reflect brain activities has become increasingly popular in driver fatigue monitoring systems \cite{9016157, 9714736}. Specifically, applying EEG can help to improve driving safety through direct and passive communication between the brain and the road environment. However, due to its nature of complexity and subject variability, EEG decoding performance in cross-subject recognition tasks is limited \cite{9508768}. In this work, we deal with the problems from the aspects of reducing the complexity of the original EEG signals by using signal decomposition methods. The original EEG signals are decomposed into components of different frequency bands. Decomposition can  facilitate the model to focus more on specific frequency bands and extract features that are more beneficial for driver fatigue recognition as there will be less impact from the less useful frequency bands. As a result, this can ease the model learning tasks from the EEG signals. }

Various decomposition methods have been applied to EEG signals to improve the decoding performance \cite{yu2022computerized}. There are mainly two categories to exploit the decomposed components. The first way is to perform hand-engineering feature extraction based on the prior knowledge of the signals, followed by classifier training. For instance, Anuragi \textit{et al.} \cite{anuragi2022epileptic} first decomposed the EEG signals by using empirical wavelet transform (EWT) \cite{gilles2013empirical}. Then, the entropy-based features were computed from the Euclidean distances of 3D phase-space representation (PSR) of the decomposed components. After the feature extraction based on the Kruskal-Wallis statistical test, ensemble learning was performed to recognize Epileptic-seizure states. Sairamya \textit{et al.} \cite{sairamya2022automatic} used relaxed local neighbor difference pattern (RLNDiP) features from the time-frequency domain, consisting of five brain rhythms obtained by discrete wavelet transform (DWT) \cite{mallat1989theory}. Artificial neural networks (ANN) were employed for the automatic diagnosis of Schizophrenia. Gu \textit{et al.} \cite{gu2021aoar} exploited non-negative matrix factorization (NMF) and empirical mode decomposition (EMD) to decompose the EEG signal into a set of intrinsic mode functions (IMFs). Statistical features were extracted from the de-noised components and inputted to classifiers such as support vector machine (SVM). PrakashYadav \textit{et al.} \cite{yadav2021variational} applied variational mode decomposition (VMD) \cite{dragomiretskiy2013variational} to decompose the EEG signals of seizure into 14 IMFs. Normalized energy, fractal dimension, number of peaks, and prominence parameters were employed as the extracted features. A Bayesian regularized shallow neural network was then proposed to perform the classification task. Interestingly, Sadiq \textit{et al.} \cite{sadiq2021exploiting} considered the decomposed components as a feature vector and further applied the methods like neighborhood component analysis (NCA) to reduce the huge dimension of the feature matrix. Although this study tried to exploit raw data, it suffered from a problem of information loss. The second category of exploiting decomposed components is that decomposition methods were exploited to reduce the noise of the signals. Feature selection was performed on each component to reduce the noise more meticulously. For instance, Aliyu \textit{et al.} \cite{aliyu2021selection} computed a series of features from the decomposed components by DWT. Then, the correlation and $p$-value analysis were used to select the optimal features. Unlike feature-wise, some other researchers directly selected the most effective components and removed the rest. For example, Wang \textit{et al.} \cite{wang2022multiband} exploited the sample entropy of each IMF decomposed by using VMD and selected the IMF of the highest sample entropy to train the classifier further. In summary, simpler patterns can be obtained after decomposition, making the signal easier to analyze. However, less attention has been paid to learning patterns directly from raw signals of decomposed components, which has the potential to extract more distinguishable patterns of the recognized states. In addition, although some feature selection methods can help reduce the noise, the problem of information loss exists simultaneously. Thus, properly capturing more distinguishable patterns while reducing the impact of redundant information from different components on model learning is still a research gap.

Furthermore, there has been an outstanding performance in exploiting deep convolutional neural networks (CNNs) to decode raw EEG signals end-to-end. Schirrmeister \textit{et al.} \cite{schirrmeister2017deep} was the first to comprehensively study the design and training of CNN by using raw EEG data. A shallow and deep CNN architecture was designed to outperform the filter-bank common spatial pattern (FBCSP) on motor imagery classification. CNNs with different architectures have been tried for various EEG-based recognition tasks. For instance, a multi-scale EEGWaveNet \cite{9645336} was proposed by Punnawish \textit{et al.} to address epileptic seizure detection. Khare \textit{et al.} \cite{9153955} exploited a CNN with the input of transformed time-frequency signals to perform emotion recognition. However, these models were only analyzed in specific domains. In 2017, Lawhern \textit{et al.} \cite{lawhern2018eegnet} proposed a compact CNN named EEGNet, which can be applied to different brain-computer interface (BCI) paradigms and achieved good performance in subject-dependent and subject-independent settings. Moreover, Cui \textit{et al.} \cite{9714736} proposed an InterpretableCNN which performed spatial and temporal convolution operations. It improved the subject-independent recognition accuracy compared with EEGNet and showed its superiority in the drowsiness detection task. Apart from the commonly used two-dimensional CNN, other architectures, such as long short-term memory (LSTM) networks \cite{wang2018lstm}, and 3D CNN \cite{zhang2021end}, have presented promising results for the corresponding BCI paradigms as well. Overall, previous studies have shown that end-to-end CNN models can improve recognition performance. However, signal complexity and uncertainty of the single model remain to be the constraints for capturing the basic patterns and achieving accurate classifications. Although decomposition could be applied together with CNN to perform recognition tasks, a single CNN may tend to overfit it and suffer from local optima when data of a higher dimension is used. Hence, exploiting CNN on raw signals of decomposed components is necessary and an ensemble model has a more robust capability of learning diverse input data compared with the single model.

To this end, we propose a novel decomposition-based ensemble CNN framework to improve the cross-subject driver fatigue recognition performance. Specifically, four decomposition methods are used to reduce signal complexity. Then, individual CNN is employed to automatically learn the beneficial patterns from each component which has a lower complexity than the original signals. Finally, two ensemble output modes are exploited to combine the classification scores before or after the Softmax layer. In addition, a component-specific batch normalization (CSBN) layer is added to reduce the subject variability of EEG signals. In the proposed framework, the diversity of the trunks in the ensemble model is provided by feeding the decomposed components with different complexity, increasing the generalization ability of the framework. We tested the framework on a challenging cross-subject driver fatigue recognition task. Results indicated that the proposed models under our framework outperformed the state-of-the-art (SOTA) methods. In particular, the DWT-based ensemble CNN boosted the performance by above 5\% as compared to the SOTA.

\textcolor{black}{The contributions of this work can be summarized as follows:}
\begin{itemize}
  \item \textcolor{black}{A novel decomposition-based ensemble CNN framework for driver fatigue recognition is proposed. 
    Using decomposition methods reduces signal complexity and enriches the input diversity by disassembling the raw EEG signals into individual components with different characteristics. Then, individual CNN is utilized to extract the beneficial features from each component. Finally, the ensemble of all the outputs of CNNs ensures strong generalizations. } 

  \item \textcolor{black}{This paper compares the use of different decomposition methods on the challenging cross-subject driver fatigue recognition task for the first time. The results demonstrated that DWT is more suitable for EEG decoding in the driver fatigue recognition task.}
  
  \item \textcolor{black}{A CSBN layer is employed to reduce subject variability.}
  
  \item \textcolor{black}{Using different architectures of CNN in the proposed framework is investigated. The framework boosts the performance of both shallow and deep CNNs. In particular, deep CNN has better recognition accuracy and is found to be more suitable for decoding decomposed components.}
\end{itemize}

\section{Preliminary study: Signal decomposition}
\label{section:decomposition_methods}
\textcolor{black}{There are two categories of signal decomposition, namely mode-based and wavelet-based methods. EMD and VMD are two representative mode-based methods. EMD \cite{huang1998empirical} is the first adaptive signal decomposition method that can analyze non-stationary and nonlinear signals. However, EMD has the  limitation of lacking mathematical theory support. VMD \cite{dragomiretskiy2013variational} has relatively stricter filter bank boundaries than EMD and avoids the cumulative error of envelope estimation caused by recursive mode decomposition. Two representative wavelet-based methods are also selected, namely EWT and DWT. EWT \cite{gilles2013empirical} also addressed the limitation of EMD which lacks mathematical formulations by building adaptive wavelet filter banks. Lastly, DWT \cite{daubechies1988orthonormal} is a classical wavelet-based method with  solid theory and is also suitable for analyzing nonlinear and non-stationary signals. These four methods are fundamental among different signal decomposition methods and have been widely used in different applications. Therefore, these four methods are selected to perform the data pre-processing in the proposed framework.} In the following parts, these four decomposition methods will be introduced in chronological order.

\subsection{Discrete wavelet transform (DWT)}
\textcolor{black}{DWT \cite{daubechies1988orthonormal} is a transform that decomposes a given signal into a number of sets, where each set is a time series of coefficients describing the time evolution of the signal in the corresponding frequency band.} Specifically, DWT is calculated by Equation \ref{eq:DWT}.
\begin{equation}
\label{eq:DWT}
\mathit{DWTf(j,k)=<x(t),\psi_{j,k} (t)>=\int x(t)\psi_{j,k}^{*} (t){\mathrm{d} t}},
\end{equation}
where $\psi_{j,k} (t)=2^{\frac{j}{2}}\psi(2^{j}t-k),j,k\in \mathbb{Z}$ is the wavelet function in DWT. In practical time series classification problems, signal $x(t)$ and $\psi_{j,k}(t)$ are both discrete as $t$ is a discrete-time index. Finite-length times series $x(t)\in L^{2}(R)$ are all applicable to DWT.

\begin{figure}[htbp]
	\centering
	\includegraphics[width=\textwidth]{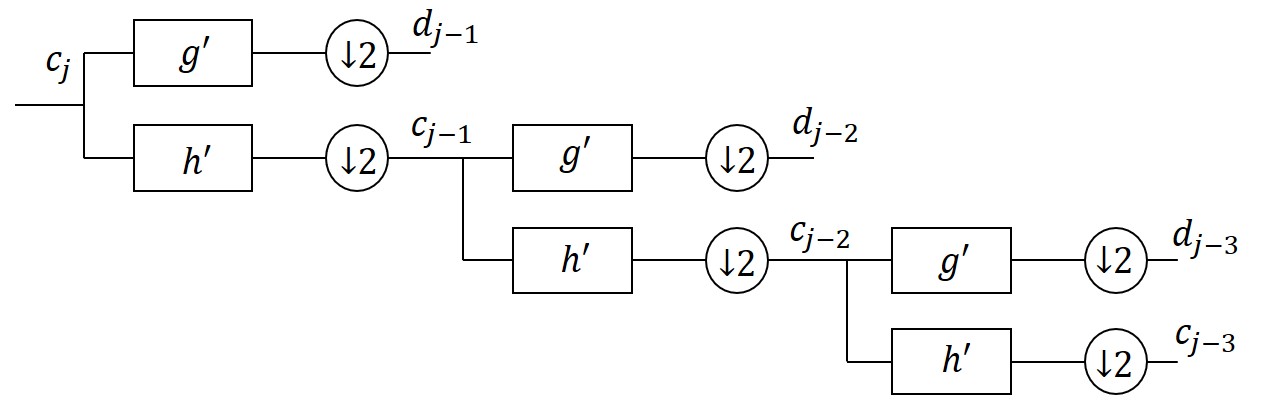}
	\caption{Decomposition of Mallat algorithm}
	\label{fig:mallat1}
\end{figure}

This paper employs the famous Mallat algorithm. In theory, approximation $A_{j}(t)$ or detail $D_{j}(t)$ in scale $j$ is calculated through the inner product between scaling function $\phi_{j,k}(t)$ or wavelet function $\psi_{j,k}(t)$ and time series $x(t)$ by Equations \ref{eq:Approx theo cal} and \ref{eq:Detail theo cal}.
\begin{equation}
\label{eq:Approx theo cal}
A_{j}(t)=\sum_{k\in
	\mathrm{Z}}<x(t),\phi_{j,k}^{*}(t)>\phi_{j,k}(t)
=\sum_{k\in\mathrm{Z}}c_{j}[k]\phi_{j,k},
\end{equation}
\begin{equation}
\label{eq:Detail theo cal}
D_{j}(t)=\sum_{k\in
	\mathrm{Z}}<x(t),\psi_{j,k}^{*}(t)>\psi_{j,k}(t)
=\sum_{k\in\mathrm{Z}}d_{j}[k]\psi_{j,k}.
\end{equation}
To get rid of the heavy computation herein, the coefficients $\left \{ \mathrm{c_{j}[k]} ,k\in \mathrm{Z}\right \}$ and  $\left \{ \mathrm{d_{j}[k]} ,k\in \mathrm{Z}\right \}$ are calculated through filtering by leveraging the nestedness of multi-resolution analysis (MRA). For the computation of $c_{j}[k]$ and $d_{j}[k]$, we can get rid of the expensive sliding inner product by replacing with the much faster convolution and down-sampling, as shown in Equation \ref{eq:downsamp conv}. 
\begin{equation}
\begin{aligned}
\label{eq:downsamp conv}
c_{j}[k] &= \sum_{i} h_{i-2k} \cdot \int f(t) \cdot 2^{(j+1)/2}\phi(2^{j+1}t-i)\mathrm{d}t\\
&=\sum_{i} h_{i-2k} \cdot c_{j+1}[i],\\
d_{j}[k] &= \sum_{i} g_{i-2k} \cdot \int f(t) \cdot 2^{(j+1)/2}\psi(2^{j+1}t-i)\mathrm{d}t\\
&=\sum_{i} g_{i-2k} \cdot c_{j+1}[I].
\end{aligned}
\end{equation}	
Viewing $\{h_{n}, n\in \mathrm{Z}\}$ and $\{g_{n}, n\in \mathrm{Z}\}$ as a pair of low-pass and high-pass filters and $c_{j+1}[i]$ as the input signal, Equation \ref{eq:downsamp conv} is implemented as shown in Figure \ref{fig:mallat1} which is the famous Mallat algorithm \cite{mallat1989theory}.

\subsection{Empirical mode decomposition (EMD)}
\textcolor{black}{EMD \cite{huang1998empirical} is a purely data-driven method of decomposing a time-domain signal into a set of different modes of oscillations and a residue. It overcomes the disadvantage of wavelet transform that a certain wavelet base needs to be selected and fixed for wavelet analysis. Since the decomposition is based on the local characteristic time scale of the data, EMD is adaptive and can be applied to nonlinear and non-stationary processes.} 

In EMD, the oscillation mode is defined as the IMF if it satisfies the following two factors \cite{huang1998empirical}: 1) The number of extremums in the oscillation and the number of zero crossings must equal or differ by at most 1; 2) The mean of the envelopes defined by the local maxima and the local minima must equal zero.

The signal decomposed by EMD can be expressed as the sum of a finite number of IMFs and a residual value shown by Equation \ref{emd_eq_first}. 
\begin{equation}
\label{emd_eq_first}
x(t)=\sum_{m=1}^k {IMF}_m(t)+r_k(t),
\end{equation}
where $k$ is the IMF number and $r_k(t)$ is the final residual value.

The set of IMFs serves as a complete, adaptive and nearly orthogonal basis for the original
signal. The algorithm of EMD is described is as follows: 
\begin{enumerate}
\item Find all the minima and maxima in $x(t)$.
\item Create an envelope of minima and maxima.
\item Find the mean $m(t)$ of the two envelopes.
\item Decrease the mean value from the signal as $d(t)=x(t)-m(t)$.
\item Check if d(t) is an IMF by applying the factors mentioned above.
\item Check if this extracted signal $d(t)$ is an IMF. If d(t) is not an IMF, iterate from step (1) to (5), considering input as d(t) to find the IMF. If d(t) is an IMF, find the residue $r(t) = x(t)-d(t)$.
\end{enumerate}

\subsection{Empirical wavelet transformation}

\textcolor{black}{EWT \cite{gilles2013empirical} is a technique that creates a multiresolution analysis of a signal using an adaptive wavelet subdivision scheme. Compared with EMD, EWT overcomes the mode mixing problem caused by the discontinuity of the time-frequency scale of the original signal. It has a complete and reliable mathematical theoretical foundation with low computational complexity.}
	
In the EWT, limited freedom is provided for selecting wavelets. The algorithm employs Littlewood-Paley and Meyer's wavelets because of the analytic accessibility of the Fourier domain's closed-form formulations \cite{spencer1994ten}. The formulation of the band-pass filters are denoted using Equations \ref{eq:EWT_1} and \ref{eq:EWT_2}.
	\begin{equation}
		\centering
		\label{eq:EWT_1}
		\hat{\phi}_{n}(\omega) = 
		\begin{cases}
			1& \makebox[1pt][r]{\text{if $\left|\omega\right| \leq (1-\gamma)\omega_{n}$}}\\
			
			\cos\left[\frac{\pi}{2}\beta\left(\frac{1}{2\gamma\omega_{n}}\left(\left|\omega\right|- (1-\gamma)\omega_{n}|\right)\right)\right]
			& \\
			\quad&\makebox[60pt][r]{\text{if $(1-\gamma)\omega_{n} \leq \left|\omega \right| \leq (1+\gamma)\omega_{n}$ }}\\
			
			0  & \text{otherwise,}\\
		\end{cases}
	\end{equation} 
	
	\begin{equation}
		\centering
		\label{eq:EWT_2}
		\hat{\psi}_{n}(\omega) = 
		\begin{cases}
			
			1& \makebox[12pt][r]{\text{if $(1+\gamma)\omega_{n} \leq \left|\omega \right| \leq (1-\gamma)\omega_{n+1}$ }}\\
			
			\cos\left[\frac{\pi}{2}\zeta\left(\frac{1}{2\gamma\omega_{n+1}}\left(\left|\omega\right|- (1-\gamma)\omega_{n+1}|\right)\right)\right]
			& \\
			\quad	&\makebox[22pt][r]{\text{if $(1-\gamma)\omega_{n+1} \leq \left|\omega \right| \leq (1+\gamma)\omega_{n+1}$ }}\\
			
			\sin\left[\frac{\pi}{2}\zeta\left(\frac{1}{2\gamma\omega_{n}}\left(\left|\omega\right|- (1-\gamma)\omega_{n}|\right)\right)\right]
			& \\
			\quad&\makebox[0.001pt][r]{\text{if $(1-\gamma)\omega_{n} \leq \left|\omega \right| \leq (1+\gamma)\omega_{n}$}}\\
			
			0  & \makebox[2pt][r]{\text{otherwise,}}\\
			
		\end{cases}
	\end{equation}

	with a transitional band width parameter $\gamma$ satisfying $\gamma \le \min_{n} \frac{\omega_{n+1}-\omega_{n}}{\omega_{n+1}+\omega_{n}}$. The most common function $\zeta(x)$ in Equation \ref{eq:EWT_1} and \ref{eq:EWT_2} is presented in Equation \ref{eq:Beta}. This empowers the formulated empirical scaling and wavelet function
	$\{\hat{\phi}_{1}(\omega),\{\hat{\psi}_{n}(\omega)\}_{n=1}^{N} \}$ to be a tight frame of $L^{2}(\mathbb{R})$ \cite{casazza2000art}.
	
	\begin{equation}
		\label{eq:Beta}
		\centering
		\beta(x)=x^{4}(35-84x+70x^2-20x^3),
	\end{equation}
	where $\{\hat{\phi}_{1}(\omega),\{\hat{\psi}_{n}(\omega)\}_{n=1}^{N} \}$ are used as the band-pass filters centered at assorted center frequencies. 
\subsection{Variational mode decomposition}
\textcolor{black}{VMD \cite{dragomiretskiy2013variational}  can adaptively find the optimal center frequency and the limited bandwidth based on the determined number of decomposed modes. VMD provides improvements over wavelet transform and EMD such that there is no modal aliasing effect and the method is not sensitive to noise. VMD can reduce the non-stationarity of time series which have high complexity and strong nonlinearity, and decompose the signals to obtain multiple IMFs with different frequency bands.}

VMD can be considered to solve the following problem as shown by Equation \ref{eq: vmd1}.
 \begin{equation}
 	\label{eq: vmd1}
 	min \left\{m_{k}\right\},\left\{w_{k}\right\}\Biggl\{\sum_{k=1}^{K}||\delta_{t}\left[(\delta(t)+\frac{k}{\pi t})\times m_{j}(t)\right]e^{k\omega_{k}t}||^{2}_{2}\Biggr\},
 \end{equation}
with the constraints of
\begin{equation}
	\sum_{k=1}^{K}m_{k}=x(t),
\end{equation}
where $m_{k}$ is mode $k$, $\omega_{k}$ is the central frequency of $m_{k}$, $K$ is the number of modes, and $x(t)$ is the input time series. The problem shown in Equation \ref{eq: vmd1} is transformed into Equation \ref{eq:6} when introducing the $L_{2}$ penalty and Lagrange multiplier.
\begin{multline}
	\label{eq:6}
	L(\left\{m_{k}\right\},\left\{w_{k}\right\},\lambda)=\alpha \Biggl\{\sum_{k=1}^{K}||\delta_{t}\left[(\delta(t)+\frac{k}{\pi t})\times m_{j}(t)\right]e^{k\omega_{k}t}||^{2}_{2}+ \\
	  ||x(t)-\sum_{k=1}^{K}m_{k}||\bigg \langle\lambda(t),x(t)- \sum_{k=1}^{K}m_{k} \rangle\Biggr\}.	
\end{multline}
The alternating direction method of multipliers (ADMM) algorithm is utilized to solve the above problem in VMD. Then, the modes $m_{k}$ and $\omega_{k}$ are obtained during the shifting process. According to the ADMM algorithm, the $m_{k}$ and $\omega_{k}$ can be computed by Equations \ref{eq: vmd2} and \ref{eq: vmd3},
\begin{equation}
\label{eq: vmd2}
	\hat{m}_{k}^{n+1}=\frac{\hat{y}(\omega)-\sum_{i\neq k}\hat{m}_{k}(\omega)+\frac{\hat{\lambda}(\omega)}{2}}{1+2\alpha(\omega-\omega_{k})^{2}}
\end{equation}
\begin{equation}
\label{eq: vmd3}
	\hat{\omega}_{k}^{n+1}=\frac{\int_{0}^{\infty}\omega|\hat{m}_{j}(\omega)|^{2}d\omega}{\int_{0}^{\infty}|\hat{m}_{j}(\omega)|^{2}d\omega},
\end{equation}
where $n$ represents the number of iterations, $\hat{y}(\omega)$, $\hat{m}_{k}(\omega)$, $\hat{\lambda}(\omega)$ and $	\hat{m}_{k}^{n+1} $ represent the Fourier transform of $x(t)$, $m_{j}(t)$, $\lambda(t)$ and $m_{k}^{n+1} $, respectively.

After decomposition, sub-band signals are obtained by EWT and DWT, as well as IMFs  are obtained by EMD and VMD. These will be represented as the term ``components" in the following sections.

\section{Methodology}
\begin{landscape}
\begin{figure*}[htbp]
  \centering
  \includegraphics[width=1.6\textwidth]{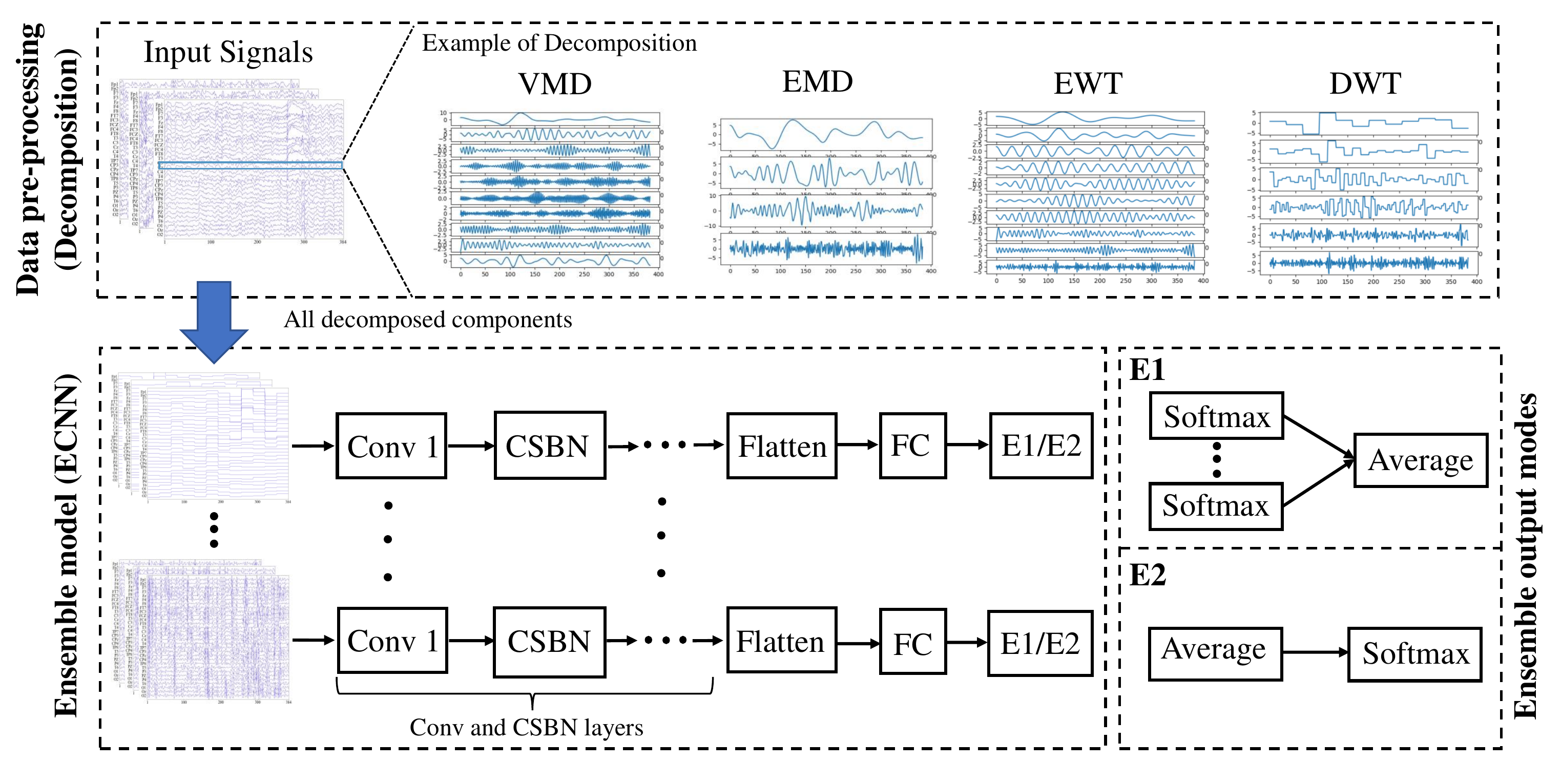}
  \caption{The structure of the decomposition-based hybrid ensemble CNN framework. Conv represents the convolutional layer.}
  \label{fig:framework}
\end{figure*}
\end{landscape}

This section presents the detailed hybrid ensemble CNN framework,  which is divided into three subsections, i.e., data pre-processing and model learning, CSBN layer and ensemble output modes. The overview of the framework is shown in Figure \ref{fig:framework}.
\subsection{Data pre-processing and model learning}
\label{Section:Data pre-processing and model learning}
Using the four decomposition methods described in Section \ref{section:decomposition_methods}, the pre-processing of EEG signals is introduced as follows. Let $\bm{X}=\{\bm{x_1}, \bm{x_2}, \dots, \bm{x_n}, \dots, \bm{x_N}\}$ be the input space (\textit{i.e.}, EEG signals) and $\bm{Y}=\{\bm{y_1}, \bm{y_2}, \dots, \bm{y_n}, \dots, \bm{y_N}\}$ be the output space (\textit{i.e.}, EEG signal categories). After applying a decomposition method with $D$ as the predefined components, the input signal is disassembled as $\bm{X_{DEC}}=\{\bm{c^1}, \bm{c^2}, \dots, \bm{c^d}, \dots, \bm{c^D}\}$, where $DEC\in \{EMD, VMD, DWT, EWT\}$ and $\bm{c_d}$ represents the $d^{th}$ disassembled component. It is worth noting that EWT and VMD methods do not have limitations on the number of decomposed components, while this parameter in EMD and DWT is associated with the nature of the data. Therefore, in EMD and DWT, when $D$ is larger than the maximum number that can be decomposed, $D$ is updated to this maximum value.

Subsequently, each raw component is fed into its corresponding CNNs of the same architecture, consisting of several convolutional layers and one fully connected (FC) layer. No specific CNN architecture is described since any CNN with adequate capability to decode raw EEG signals can be exploited as a trunk in our framework. Only the feed-forward propagation and the parameter updating process are introduced here. Specifically, supposing that there are $L$ convolutional layers in the model, the hidden features for component $d$ in layer $l$ are denoted as $\bm{h^d_l}\in \mathbb{R}^{N\times C_l\times H^l\times W^l}$, where $C_l$ is the number of channels produced by the convolution and $H^l\times W^l$ is the size of the 2D feature map. After propagating the input through the convolutional and CSBN layers, which will be further introduced in Section \ref{section:CSBN}, the final feature map $\bm{h^d_L}$ is flattened and fed into a FC layer that transforms the features into two classification scores: $\bm{s^d_1}$ and $\bm{s^d_2}$, corresponding to two classes, alert and fatigue, respectively. Finally, two ensemble output modes (E1 and E2), which will be described in detail in Section \ref{section:Ensemble output modes}, are used to integrate the classification scores. Lastly, the CNN of each component is combined as an ensemble model, which is denoted as ``ECNN". 

Overall, in the proposed framework, there are four decomposition methods and two ensemble output modes, which give rise to eight models under our framework as listed in Table \ref{table:eight models}. The framework can be extended to any backbone CNN model that has adequate capability to decode EEG signals.

\begin{table}[!h]
\centering
\caption{The eight proposed models under the decomposition-based hybrid ensemble CNN framework}
\label{table:eight models}
\resizebox{\textwidth}{!}{
\begin{tabular}{c|cccc}
\hline
   & VMD        & EMD        & EWT        & DWT        \\ \hline
E1 & VMD+ECNN(E1) & EMD+ECNN(E1) & EWT+ECNN(E1) & DWT+ECNN(E1) \\ 
E2 & VMD+ECNN(E2) & EMD+ECNN(E2) & EWT+ECNN(E2) & DWT+ECNN(E2) \\ \hline
\end{tabular}}
\end{table}

\subsection{Component specific batch normalization (CSBN) layer}
\label{section:CSBN}
In the framework, a CSBN layer is employed to decrease subject variability. Batch normalization (BN) \cite{ioffe2015batch} is usually exploited to alleviate the issue of internal covariate shifting. Standardizing hidden features is first performed in a standard BN layer, and then two affine parameters $\gamma$ and $\beta$ are used to transform the inherent mean and variance into trainable variables. Therefore, for a channel of hidden features, $\bm{h^d_{l,c}}$, the operation of BN is expressed as:
\begin{equation}
\label{eq:bn+layer}
BN(\bm{h_{l,c}^d})=\gamma\bm{\hat{h}_{l,c}^d}+\beta,
\end{equation}
where $c$ represents the $c^{th}$ channel of hidden features and $\bm{\hat{h}_{l,c}^d}$ is the result of standardizing the hidden features, which is given by:
\begin{equation}
\label{eq:standardization}
\bm{\hat{h}_{l,c}^d}=\frac{\bm{h_{l,c}^d}-\mu}{\sqrt{\sigma^2+\epsilon}},
\end{equation}
where $\epsilon>0$. \textcolor{black}{$\mu$ and $\sigma^2$ are the mean and variance of a channel of hidden features within a mini-batch, which are computed by}
\begin{align}
\label{eq:mean}
\mu&=\frac{1}{K\times H^l\times W^l}\sum_{k, h, w}\bm{h^d_{l,c}}[k, h, w], \\
\label{eq:std}
\sigma^2&=\frac{1}{K\times H^l\times W^l}\sum_{k, h, w}(\bm{h^d_{l,c}}[k, h, w]-\mu)^2,
\end{align}
\textcolor{black}{where $k\in [1,K]$, $h\in[1, H^l]$, $w\in[1, W^l]$ and $K$ represents the number of samples in a mini-batch. } 

While BN exploits the statistics (\textit{i.e.}, $\mu$ and $\sigma^2$) of the training data to standardize the testing data, there is a problem of different distribution between the training and testing sets. This can be regarded as a domain shift problem, where the data of the training subjects and the data of the testing subjects are considered as the source domain and the target domain, respectively. Hence, adaptive BN (AdaBN) \cite{li2018adaptive} was proposed to reduce the domain shift. \textcolor{black}{Specifically, unlike the BN layer, of which the mean and the standard deviation used for testing are directly adopted from the training data, the AdaBN layer adopts domain specific normalization
for different domains.}

\textcolor{black}{Therefore, as we employ AdaBN on the hidden features of each component in the proposed ensemble model, we denote this component-specific AdaBN as CSBN. This is to say that for the model of each component in the proposed ensemble model, the mean and the standard deviation of the hidden features used by the CSBN layer are automatically calculated during testing.} Based on the CSBN, the hidden features of both source and target domains can be aligned in different complexity levels, increasing the model's generalization ability on any unseen subject data. The operation of the CSBN layer in the source or target domain is described as:
\begin{equation}
\label{eq:CSBN}
CSBN(\bm{h_{t}^{d,l,c}})=\gamma\bm{\hat{h}_{t}^{d,l,c}}+\beta,
\end{equation}
where $t$ represents the source or target domain. \textcolor{black}{For convenience, $h^d_{l,c}$ is rewritten as $h^{d,l,c}$ here. Then, Equation \ref{eq:standardization}-\ref{eq:std} can be reformulated as}
\begin{gather}
\label{eq:CSBN standardization}
\bm{\hat{h}_{t}^{d,l,c}}=\frac{\bm{h_{t}^{d,l,c}}-\mu_t}{\sqrt{\sigma_t^2+\epsilon}}, \\
\label{eq:CSBN mean}
\mu_t=\frac{1}{K\times H_t^l\times W_t^l}\sum_{k, h, w}\bm{h^{d,l,c}_{t}}, \\
\label{eq:CSBN std}
\sigma_t^2=\frac{1}{K\times H_t^l\times W_t^l}\sum_{k, h, w}(\bm{h^{d,l,c}_{t}}-\mu_t)^2.
\end{gather}

\subsection{Ensemble output modes}
\label{section:Ensemble output modes}
Two ensemble modes, E1 and E2, are employed to integrate the classification scores $\bm{s^d_1}$ and $\bm{s^d_2}$ in different trunks of the ensemble model, which are described as follows.

1) E1: To force the automatic learning of the proper patterns for integration, the parameters of the models in different trunks are updated simultaneously. Specifically, the output scores are averaged by
\begin{equation}
\label{eq:average_scores}
\bm{s_i}=\frac{1}{D}\sum_{d=1}^{D} \bm{s^d_i},
\end{equation}
where $i$ represents the index of the score. 

Then, the average score is transformed to the probabilities by the Softmax layer, which is described as:
\begin{equation}
\label{eq:Softmax}
\bm{p_i}=\frac{exp(\bm{s_i})}{\sum_j{exp(\bm{s_j})}}.
\end{equation}
Finally, denoting $F(\cdot)$ as the output of a forward pass in the integrated model, the parameters in separate trunks are updated together in back-propagation by using the cross-entropy loss:
\begin{equation}
\label{eq:cross_entropy}
\mathcal{L}=\mathbb{E}_{(x_n, y_n)\in (X, Y)}-\sum_{n=1}^{N}y_nlogF(\bm{x_n}).
\end{equation}

2) E2: Based on the diversity of the classification scores obtained by training CNN models with different components, the second ensemble mode (E2) employs majority voting to directly ensemble the output probabilities of the trunks. Different from the first ensemble mode (E1), each trunk has an individual Softmax layer. Consequently, the specific output probabilities for each input component are calculated as $\bm{p^d_i}$ based on Equation \ref{eq:Softmax}. Moreover, the individual model in each trunk are trained separately by exploiting Equation \ref{eq:cross_entropy}. After updating the model parameter, the final output probabilities are given by
\begin{equation}
\label{eq:majority_voting}
\bm{p_i}=\frac{1}{D}\sum_{d=1}^{D}\bm{p_i^d},
\end{equation}
which is known as the soft voting. Finally, the predicted class can be obtained based on the output probabilities.

\section{Experiments}
\label{sec:results}
\subsection{Introduction of public driving dataset}
The public driving dataset recorded from a sustained-attention driving task \cite{cao2019multi} was used in this work. In the experiment, lane-departure events were randomly induced to make the car drift from the original cruising lane towards the left or right side (deviation onset). Each participant was instructed to quickly compensate for this perturbation by steering the wheel (response onset) to cause the car to move back to the original cruising lane (response offset). A complete trial included the deviation onset, response onset, and response offset events. \textcolor{black}{During the experiment, the EEG activity of each subject was recorded  using a 32-channel Quik-Cap (Compumedics NeuroScan Inc.) following the International 10-20 system of electrode placement with Ag/AgCl electrodes. In this work, processed data \cite{cao2019multi} were used in our study. The pre-processing steps included bandpass filtering and artefact rejection. The bandpass finite impulse response filters of 1-50 Hz were applied to remove low-frequency direct current drifts and power line noise. For artefact rejection, the apparent eye blink contamination in the EEG signals was manually removed by visual inspection. Following that, artefacts were removed by the Automatic Artifact Removal plug-in for EEGLAB, which provided automatic correction of ocular and muscular artefacts in the EEG signals. }

\textcolor{black}{Regarding sample extraction, in practice, short-interval EEG data prior to the deviation onset can be adopted to perform fatigue recognition. In this work, to evaluate the proposed framework and to perform a fair comparison with the SOTA works, the 3-second EEG data prior to the deviation onset which was commonly exploited in previous methods \cite{wei2018toward} \cite{9714736} was adopted to classify the `alert’ vs. `fatigue’ on the upcoming lane-departure event. }

\textcolor{black}{Then, we followed Wei \textit{et al.} \cite{wei2018toward}  to adopt the local reaction time (RT) and the global RT to label data. Specifically, the RT was defined as the time between the deviation onset and the response onset. For each subject, the RT in each lane-departure event was named as the local RT, which represented the short-term level of fatigue. On the other hand, the long-term level of fatigue was defined by the global RT, which was calculated by averaging the RTs across all trials within a 90-second window before the deviation onset. The `alert-RT' was calculated for each driving session as the $5^{th}$ percentile of the local RTs, which represented the RT that the subject could execute while being alert.} When both the local and global RT was shorter than 1.5 times the alert-RT, the corresponding extracted EEG data was labeled as `alert'. When both the local and global RT were longer than 2.5 times the alert-RT, the data was labeled as `fatigue'. Transitional states with moderate performance were excluded, and the neutral state was not considered in this work. For the subjects with multiple datasets, we selected the most balanced one to perform the filter operation. Then, we further down-sampled the data to 128 Hz. Finally, we obtained a balanced driver fatigue dataset which included 2022 samples of 11 subjects. \textcolor{black}{The specific number of samples for each subject was shown in Table \ref{table:Number of samples for each subject}.} The data size of one sample is 30 (channels) $\times$ 384 (sample points).

\begin{table}[]\tiny
\centering
\caption{Number of samples for each subject}
\label{table:Number of samples for each subject}
\resizebox{0.5\textwidth}{!}{
\begin{tabular}{ccc}
\hline
\multirow{2}{*}{Subject ID} & \multicolumn{2}{c}{Number of samples} \\ \cline{2-3} 
                            & Alert            & Fatigue            \\ \hline
1                           & 94               & 94                 \\
2                           & 66               & 66                 \\
3                           & 75               & 75                 \\
4                           & 74               & 74                 \\
5                           & 112              & 112                \\
6                           & 83               & 83                 \\
7                           & 51               & 51                 \\
8                           & 132              & 132                \\
9                           & 157              & 157                \\
10                          & 54               & 54                 \\
11                          & 113              & 113                \\
Total                       & 1011             & 1011               \\ \hline
\end{tabular}}
\end{table}

\subsection{Cross-subject driver fatigue recognition results}
\label{section:Cross-subject situation awareness recognition results}
\subsubsection{Experiment Settings}
\label{section:Experiment Settings}
The experiment was conducted on an Alienware Desktop with a 64-bit Windows 10 operation system powered by Intel(R) Core(TM) i7-6700 CPU and a NVIDIA GeForce GTX 1080 graphics card. The codes were implemented and tested on the platform of Python 3.7.0. Pytorch framework was employed in this work.

Considering the high complexity of EEG signals, we first set the number of decomposed components as 10. The specific sensitivity test of the number of components will be described in Section \ref{section:Sensitivity test of four decomposition methods}. After the decomposition, the highest number of decomposed components that could be obtained from EMD and DWT was 4 and 6. respectively. Finally, the number of components used in the following evaluation for VMD, EMD, EWT and DWT are 10, 4, 10 and 6, respectively.

To validate the effectiveness of the decomposition-based hybrid ensemble CNN framework, the eight proposed models under the framework as previously described in Section \ref{Section:Data pre-processing and model learning} was compared with strong baselines in the setting of leave-one-subject-out (LOSO) cross-validation (CV). \textcolor{black}{Since the proposed framework involves the use of the CNN model and the transfer learning technique which is the use of the CSBN layer, the proposed methods were compared with two categories of methods to demonstrate the superiority of the proposed framework in cross-subject driver fatigue recognition. Specifically, the first category of methods for comparison included the support vector machine (SVM) classifier \cite{wei2018toward} with the input of extracted power spectral density (PSD) features, EEGNet \cite{lawhern2018eegnet}, ShallowCNN \cite{schirrmeister2017deep} and InterpretableCNN (ICNN) \cite{9714736}. Among them, SVM+PSD represented the methods that were based on the hand-engineering features, while EEGNet and ShallowCNN were the commonly used end-to-end EEG decoding baselines with the input of raw EEG data. Lastly, ICNN was the SOTA method on the used public driving task. ICNN was also the end-to-end model with the input of raw EEG data. Furthermore, the second category of methods included the SOTA transfer learning techniques for driver fatigue recognition which are TCA \cite{liu2020inter} and MIDA \cite{liu2020inter}. These two methods used the extracted PSD features as the inputs. The SM model \cite{LI2022136} which performed end-to-end EEG decoding was also included for comparison. To demonstrate the effectiveness of the proposed framework in the driving context, all baseline methods mentioned were compared on the same extracted data.}

The ICNN, the SOTA model, was first employed as the backbone network of the proposed framework to compare with the baseline methods. The use of different backbone networks will be compared and discussed in Section \ref{section:Investigation on different backbone networks}.

The PSD features used in this work were computed via Fast Fourier Transform on each EEG epoch from these four spectral bands: delta (1–4 Hz), theta (4–8 Hz), alpha (8–12 Hz) and beta (12–30 Hz). The final feature vector was a concatenation of spectral powers extracted from the four bands and all available channels. In this study, the final feature vector was of $4$ (frequency bands) $\times$ $30$ (channels) $=\ 120$ dimensions. 

Regarding the comparison among the back-prop-based models, all hyper-parameters were set as the same. Specifically, Adam optimizer was set as momentum $\beta_1=0.9$ and $\beta_2=0.99$. The mini-batch was 50, and the learning rate was 0.001. The model was trained for 50 epochs.

\subsubsection{Comparison results}
The LOSO average accuracy (denoted as avg. acc. in tables) of the eight proposed ensemble models under the decomposition-based hybrid ensemble CNN framework and the baseline algorithms in the cross-subject driver fatigue recognition task was compared in Table \ref{comparison_table_1}. Overall, we can observe that the eight proposed models under the framework all outperformed the baseline methods. Compared with the SOTA results of ICNN, the increase in LOSO average accuracy ranged from 0.15\% to 3.76\%. In particular, the DWT-based ensemble CNN in E2 mode (DWT+EICNN(E2)) achieved the best LOSO average accuracy of 82.11\% amongst the proposed models. \textcolor{black}{Better performance of the proposed models demonstrated the effectiveness of the framework on fatigue recognition in the driving context.} 

\begin{table*}[htbp]
\caption{Comparison between the LOSO average accuracy (\%) of decomposition-based hybrid ensemble CNN framework and the baseline methods. EICNN: ensemble ICNN.}
\label{comparison_table_1}
\resizebox{\textwidth}{!}{
\begin{tabular}{c|ccccccccccc|c}
\hline
\multirow{2}{*}{Methods}                                 & \multicolumn{11}{c|}{Subjects}                                                                                                                                                           & \multirow{2}{*}{Avg. Acc.} \\ \cline{2-12}
                                                         & 1              & 2              & 3              & 4              & 5              & 6              & 7              & 8              & 9              & 10             & 11             &                             \\ \hline
SVM \cite{wei2018toward}                                                 & 77.66          & 75.76          & 66.67          & 66.22          & 83.04          & 75.90          & 59.80          & 67.80          & 88.54          & 70.37          & 59.73          & 71.95                       \\
TCA \cite{liu2020inter}                 & 90.43          & 53.03          & 70.67          & 72.97          & 79.91          & 78.31          & 63.73          & 68.94          & 81.85          & 83.33          & 61.50          & 73.15                       \\
MIDA \cite{liu2020inter}                & 80.32          & 59.09          & 80.67          & 77.03          & 82.14          & 76.51          & 51.96          & 71.97          & 87.26          & \textbf{85.19} & 55.31          & 73.40                       \\
ShallowCNN \cite{schirrmeister2017deep} & 66.49          & 45.45          & 77.33          & 70.95          & 86.16          & 83.73          & \textbf{70.59} & \textbf{81.44} & 81.21          & 75.93          & 75.66          & 74.09                       \\
EEGNet \cite{lawhern2018eegnet}     & 76.06          & 66.67          & 61.33          & 81.76          & 78.13          & 75.90          & 66.67          & 75.38          & 78.66          & \textbf{85.19} & 69.47          & 74.11                       \\
SM model \cite{LI2022136}               & 78.72          & 68.18          & 79.33          & 68.24          & 85.27          & 83.73          & 64.71          & 57.2           & 78.03          & 82.41          & 71.68          & 74.32                       \\
ICNN \cite{9714736}                     & 85.00          & 67.65          & 81.80          & 78.99          & 88.35          & 83.92          & 67.06          & 79.05          & 89.17          & 71.02          & 69.82          & 78.35                       \\ \hline
VMD+EICNN(E1)                                             & 89.89          & 78.03          & 75.33          & 71.62          & 86.61          & 84.34          & 68.63          & 74.62          & 87.26          & 75.93          & 71.24          & 78.50                       \\
EWT+EICNN(E1)                                             & 88.83          & 80.30          & 77.33          & 77.03          & 88.84          & 80.12          & 68.63          & 71.21          & 88.85          & 77.78          & 71.68          & 79.15                       \\
EMD+EICNN(E1)                                              & 89.89          & 72.73          & 81.33          & 80.41          & 87.95          & 84.34          & 67.65          & 77.27          & 89.17          & 80.56          & 71.68          & 80.27                       \\
DWT+EICNN(E1)                                              & 88.83          & 73.48          & 82.67          & 81.08          & \textbf{91.96} & 86.14          & 67.65          & 80.30          & 89.17          & 76.85          & \textbf{79.20} & 81.58                       \\
VMD+EICNN(E2)                                              & 89.89          & 75.76          & 71.33          & 71.62          & 86.61          & 83.73          & 67.65          & 75.00          & 88.54          & 83.33          & 75.22          & 78.97                       \\
EWT+EICNN(E2)                                            & 88.30          & 81.82          & 77.33          & 79.05          & 89.29          & 84.94          & 66.67          & 77.27          & \textbf{89.81} & 79.63          & 76.99          & 81.01                       \\
EMD+EICNN(E2)                                             & \textbf{91.49} & \textbf{87.88} & 80.00          & 75.68          & 89.29          & 81.93          & 62.75          & 81.06          & 87.58          & 81.48          & 74.34          & 81.22                       \\
DWT+EICNN(E2)                                             & 88.30          & 73.48          & \textbf{83.33} & \textbf{82.43} & 90.63          & \textbf{90.96} & 64.71          & \textbf{81.44} & 88.54          & 81.48          & 77.88          & \textbf{82.11}              \\ \hline
\end{tabular}}
\end{table*}

As for the classification performance for the individual subject data, the decomposition-based hybrid ensemble models also presented the highest accuracy among all compared methods for all subjects except for subject 7 and 10. This could be due to the much lower decoding capability of the backbone network--ICNN in nature, which was shown from the lower accuracy results for subject 7 and 10 as obtained by ICNN. 

\textcolor{black}{To better understand the capability of the proposed framework for driver fatigue classification, Precision, Sensitivity, Specificity and F1-score of the proposed models and the baseline methods were compared in Table \ref{Precision, Sensitivity and F1-score}. The class `alert' was set as positive, while the class `fatigue' was set as negative in the calculation of these metrics. It was observed that all eight proposed models under the framework showed better Precision than that of the baseline methods. Most of them presented better F1-score than the baseline methods except for the VMD-based ensemble CNN in both E1 and E2 modes and the EWT-based ensemble CNN in E1 mode. It is worth noting that the DWT-based ensemble CNN in both modes achieved the highest F1-score among all compared methods.} \textcolor{black} {Furthermore, the Specificity of all the proposed models was higher than that of the baseline methods. This demonstrated that the proposed models could better ensure that those who are fatigued are indeed classified as `fatigue'. This is extremely meaningful in practice. Lastly, the proposed models showed higher Sensitivity than most of the baseline methods.} 

\begin{table}[]
\centering
\caption{Precision, Sensitivity, Specificity and F1-score (\%) of decomposition-based hybrid ensemble CNN framework and the baseline methods. EICNN: ensemble ICNN.}
\label{Precision, Sensitivity and F1-score}
\resizebox{0.6\textwidth}{!}{
\begin{tabular}{c|cccc}
\hline
             & Precise        & Sensitivity    & Specificity    & F1-Score       \\ \hline
SVM \cite{wei2018toward}      & 73.76          & 73.39          & 73.89          & 73.57          \\
TCA \cite{liu2020inter}           & 74.35          & 73.39          & 74.68          & 73.87          \\
MIDA \cite{liu2020inter}          & 73.96          & 75.57          & 73.39          & 74.76          \\
EEGNet \cite{lawhern2018eegnet}       & 69.04          & \textbf{89.12} & 60.04          & 77.81          \\
ShallowCNN \cite{schirrmeister2017deep}      & 72.74          & 82.59          & 69.04          & 77.35          \\
SM model \cite{LI2022136}     & 70.66          & 82.89          & 65.58          & 76.28          \\
ICNN \cite{9714736}          & 77.78          & 83.78          & 76.06          & 80.67          \\
VMD+EICNN+E1 & 78.05          & 82.29          & 76.85          & 80.12          \\
EWT+EICNN+E1 & 78.50          & 82.69          & 77.35          & 80.54          \\
EMD+EICNN+E1 & 80.08          & 83.48          & 79.23          & 81.74          \\
DWT+EICNN+E1 & 81.77          & 85.16          & 81.01          & 83.43          \\
VMD+EICNN+E2 & 78.48          & 82.99          & 77.25          & 80.67          \\
EWT+EICNN+E2 & 80.51          & 84.97          & 79.43          & 82.68          \\
EMD+EICNN+E2 & 82.02          & 84.37          & \textbf{81.50} & 83.18          \\
DWT+EICNN+E2 & \textbf{82.07} & 85.56          & 81.31          & \textbf{83.78} \\ \hline
\end{tabular}}
\end{table}

We further discuss the effect of the two ensemble modes and the four decomposition methods in the framework.  

1) \textbf{Two ensemble modes:} From Table \ref{comparison_table_1}, for a specific decomposition method-based ensemble model, a higher LOSO average accuracy can be observed when using E2 mode as compared to using E1 mode. One possible reason is that during each simultaneous update in E1 mode, the low-quality patterns obtained may disturb the feature learning of the proper components for classification. Since the E2 mode performed better than the E1 mode, the ablation study in Section \ref{Ablation} was mainly investigated in the E2 mode.

2) \textbf{Four decomposition methods:} With the same parameter of 10 decomposed components for VMD and EWT and in the same E2 mode, the ensemble model based on EWT presented better results of LOSO average accuracy. Notably, the EMD- and DWT-based ensemble models in E2 mode obtained an even higher LOSO average accuracy than that of VMD and EWT. This could be because that the signals were decomposed into the simplest components in EMD and DWT. Hence, the models can capture beneficial features more easily.

To demonstrate the superiority of the proposed models, the one-tailed Wilcoxon paired signed-rank test was conducted as shown in Table \ref{Wilcoxon}. Only the $p$-values with significance ($p<0.05$) were listed in the table. We can observe that the DWT-based models in E2 mode consistently showed a significantly stronger capability of classification than all baseline methods, while other proposed models showed significant improvements in LOSO average accuracy over most of the baselines.

\begin{table*}[]
\caption{Wilcoxon test results of the proposed models and the baseline methods. VMD(E1): VMD+EICNN(E1); EWT(E1): EWT+EICNN(E1); EMD(E1): EMD+EICNN(E1); VMD(E2): VMD+EICNN(E2).}
\label{Wilcoxon}
\centering
\resizebox{\textwidth}{!}{
\begin{tabular}{c|ccccccccccc}
\hline
               & SVM \cite{wei2018toward} & TCA \cite{liu2020inter}   & MIDA \cite{liu2020inter}   & ShallowCNN \cite{schirrmeister2017deep} & EEGNet 8,2 \cite{lawhern2018eegnet} & SM model \cite{LI2022136} & ICNN \cite{9714736}   & VMD(E1) & EWT(E1) & EMD(E1) & VMD(E2) \\ \hline
VMD+EICNN(E1)  & 0.0010  & 0.0336 &        &            &            &          &        &         &                &                &               \\
EWT+EICNN(E1) & 0.0005  & 0.0092 & 0.0372 &            &            & 0.0297   &        &         &                &                &               \\
EMD+EICNN(E1) & 0.0015  & 0.0024 & 0.0034 & 0.0415     & 0.0093     & 0.0047   & 0.0372 &         &                &                &               \\
DWT+EICNN(E1) & 0.0015  & 0.0034 & 0.0048 & 0.0068     & 0.0068     & 0.0049   & 0.0025 & 0.0269  &                &                &               \\
VMD+EICNN((E2) & 0.0038  & 0.0083 &        &            & 0.0508     & 0.0183   &        &         &                &                &               \\
EWT+EICNN(E2)  & 0.0005  & 0.0034 & 0.0122 & 0.0463     & 0.0109     & 0.0093   &        & 0.0049  & 0.0234         &                & 0.0415        \\
EMD+EICNN(E2)  & 0.0010  & 0.0034 & 0.0161 &            & 0.0161     & 0.0161   &        & 0.0337  & 0.0508         &                &               \\
DWT+EICNN(E2)  & 0.0035  & 0.0049 & 0.0024 & 0.0083     & 0.0049     & 0.0035   & 0.0049 & 0.0210  &                & 0.0415         & 0.0463        \\ \hline
\end{tabular}}
\end{table*}

According to the above results, the proposed decomposition-based hybrid ensemble CNN framework showed significantly superior performance on the challenging cross-subject driver fatigue recognition task as compared to the baseline methods. Among the proposed models under the framework, the DWT-based ensemble ICNN model in E2 mode showed the highest LOSO average accuracy with a significant difference from the others. Therefore, the proposed framework is effective in boosting the performance of backbone models on the EEG-based cross-subject driver fatigue recognition task.

\subsection{Ablation study}
\label{Ablation}

\subsubsection{Sensitivity test of four decomposition methods}
\label{section:Sensitivity test of four decomposition methods}

After decomposition, the total number of decomposed components fed into the ensemble models will affect the classification performance. In Section \ref{section:Cross-subject situation awareness recognition results}, we first set the total number of decomposed components fed into the ensemble models as 10, 4, 10, and 6 for the four decomposition methods, VMD, EMD, EWT, and DWT, respectively. The total number of components in the range of 3 to 10 were then selected for further investigation using the ensemble ICNN in E2 mode. The LOSO average accuracy of using the different total number of components for training was listed in Table \ref{Sensitive_test}. For EMD and DWT, since the maximum number of decomposed components that can be obtained was 4 and 6, respectively, the number of components below this maximum was investigated. 

The results showed the best performance in the VMD- and DWT-based models trained with 5 decomposed components. For the EMD- and EWT-based models, using 4 and 9 decomposed components, respectively, obtained better accuracy. In general, for the EWT-, EMD- and EWT-based models, an overall increasing trend of LOSO average accuracy could be observed with higher total number of components. This could be because that as the total number of decomposed components for training increased, the complexity of each decomposed component could be reduced. This may ease the model learning and lead to better recognition performance. \textcolor{black}{However, the VMD-based model did not show a similar trend. Large fluctuations were observed. This may be attributed to the characteristic of VMD. According to the original work of VMD \cite{dragomiretskiy2013variational}, the performance of VMD may be compromised when it comes to non-stationary signals such as long-term EEG signals, as drastic change and global overlapping may occur in the spectral bands of modes. In other words, possibly due to the impact of the nature of EEG signals on VMD, the performance of the VMD-based model may be very sensitive to the total number of decomposed components. This may explain why the VMD-based model only showed high LOSO average accuracy at certain total numbers of components.}

\begin{table}[!h]
\centering
\caption{LOSO average accuracy (\%) of using different total number of decomposed components for training.}
\label{Sensitive_test}
\resizebox{0.7\textwidth}{!}{
\begin{tabular}{c|cccccccc}
\hline
\multirow{2}{*}{Methods} & \multicolumn{8}{c}{Number of components}                      \\ \cline{2-9} 
                         & 3     & 4     & 5     & 6     & 7     & 8     & 9     & 10    \\ \hline
VMD                      & 79.28 & 80.54 & 80.92 & 78.90 & 79.36 & 78.84 & 79.60 & 78.97 \\
EWT                      & 79.47 & 80.06 & 79.59 & 80.74 & 80.38 & 80.81 & 81.05 & 81.01 \\
EMD                      & 80.66 & 81.22       &       &       &       &       &       &       \\
DWT                      & 81.32 & 81.46 & 82.15 & 82.11  &       &       &       &       \\ \hline
\end{tabular}}
\end{table}

\subsubsection{Investigation on the use of different backbone networks in proposed framework}
\label{section:Investigation on different backbone networks}

To demonstrate that any backbone CNN with adequate decoding capability can be utilized in the proposed framework, the comparison results of using two different networks: a deep CNN (EEGNet-8,2 \cite{lawhern2018eegnet}) and a shallow CNN (ShallowCNN \cite{schirrmeister2017deep}) as the backbone was presented in Table \ref{Backbone_Nets}. Since the original version of ShallowCNN does not include batch normalization (BN) layers, we adopted a variant with BN layers to the outputs of the convolutional layers' output. The number of decomposition levels for each decomposition was the same as the settings described in Section \ref{section:Experiment Settings}. Results showed that both decomposition-based ensemble ShallowCNN in E2 mode and decomposition-based ensemble EEGNet in E2 mode achieved competitive performance compared with the corresponding baselines. In particular, the ensemble EEGNet showed better performance than the ensemble ShallowCNN, which may be attributed to the better decoding capability of the deep CNN on high-complexity components.

\begin{table*}[]
\centering
\caption{LOSO average accuracy (\%) of using different backbone networks in proposed framework. EShallowCNN: ensemble ShallowCNN; EEEGNet: ensemble EEGNet.}
\label{Backbone_Nets}
\resizebox{\textwidth}{!}{
\begin{tabular}{c|ccccccccccc|c}
\hline
\multirow{2}{*}{Model} & \multicolumn{11}{c|}{Subjects}                                                                                                                                                           & \multirow{2}{*}{Avg. Acc.} \\ \cline{2-12}
                       & 1              & 2              & 3              & 4              & 5              & 6              & 7              & 8              & 9              & 10             & 11             &                             \\ \hline
Baseline-ShallowCNN \cite{schirrmeister2017deep}    & 66.49          & 45.45          & 77.33          & 70.95          & 86.16          & 83.73          & \textbf{70.59} & 81.44          & 81.21          & 75.93          & 75.66          & 74.09                       \\
VMD+EShallowCNN(E2)    & 82.98          & 44.70          & 72.00          & 79.05          & 83.93          & 89.76          & 68.63          & 72.35          & \textbf{90.13} & 78.70          & \textbf{83.19} & 76.86                       \\
EWT+EShallowCNN(E2)    & 84.57          & 80.30          & 74.67          & 70.27          & 88.84          & 83.73          & 66.67          & 75.76          & \textbf{90.13} & 75.00          & 79.20          & 79.01                       \\
EMD+EShallowCNN(E2)    & 76.06          & 53.79          & 80.00          & 79.05          & 87.95          & 85.54          & 67.65          & 78.03          & 87.90          & 82.41          & 78.32          & 77.88                       \\
DWT+EShallowCNN(E2)    & 86.70          & 57.58          & \textbf{83.33} & \textbf{85.14} & 90.18          & 88.55          & \textbf{70.59} & \textbf{84.09} & 86.94          & 78.70          & 81.42          & 81.20                       \\ \hline
Baseline-EEGNet 8,2 \cite{lawhern2018eegnet}    & 76.06          & 66.67          & 61.33          & 81.76          & 78.13          & 75.90          & 66.67          & 75.38          & 78.66          & 85.19          & 69.47          & 74.11                       \\
VMD+EEEGNet(E2)        & 90.43          & 71.97          & 72.00          & 74.32          & 88.39          & 84.34          & 69.61          & 72.35          & 87.26          & \textbf{86.11} & 75.22          & 79.27                       \\
EWT+EEEGNet(E2)        & 88.30          & 80.30          & 78.67          & 79.05          & 87.95          & 88.55          & 63.73          & 73.86          & 89.17          & 81.48          & 79.65          & 80.97                       \\
EMD+EEEGNet(E2)        & \textbf{90.96} & \textbf{84.85} & 81.33          & 78.38          & 91.07          & 92.17          & 68.63          & 77.65          & \textbf{90.13} & 79.63          & 73.89          & 82.61                       \\
DWT+EEEGNet(E2)        & 89.36          & 72.73          & \textbf{83.33} & 80.41          & \textbf{94.20} & \textbf{95.78} & 68.63          & 83.33          & 87.58          & 83.33          & 79.65          & \textbf{83.48}              \\ \hline
\end{tabular}}
\end{table*}

\begin{figure}[htbp]
  \centering
  \includegraphics[width=0.7\textwidth]{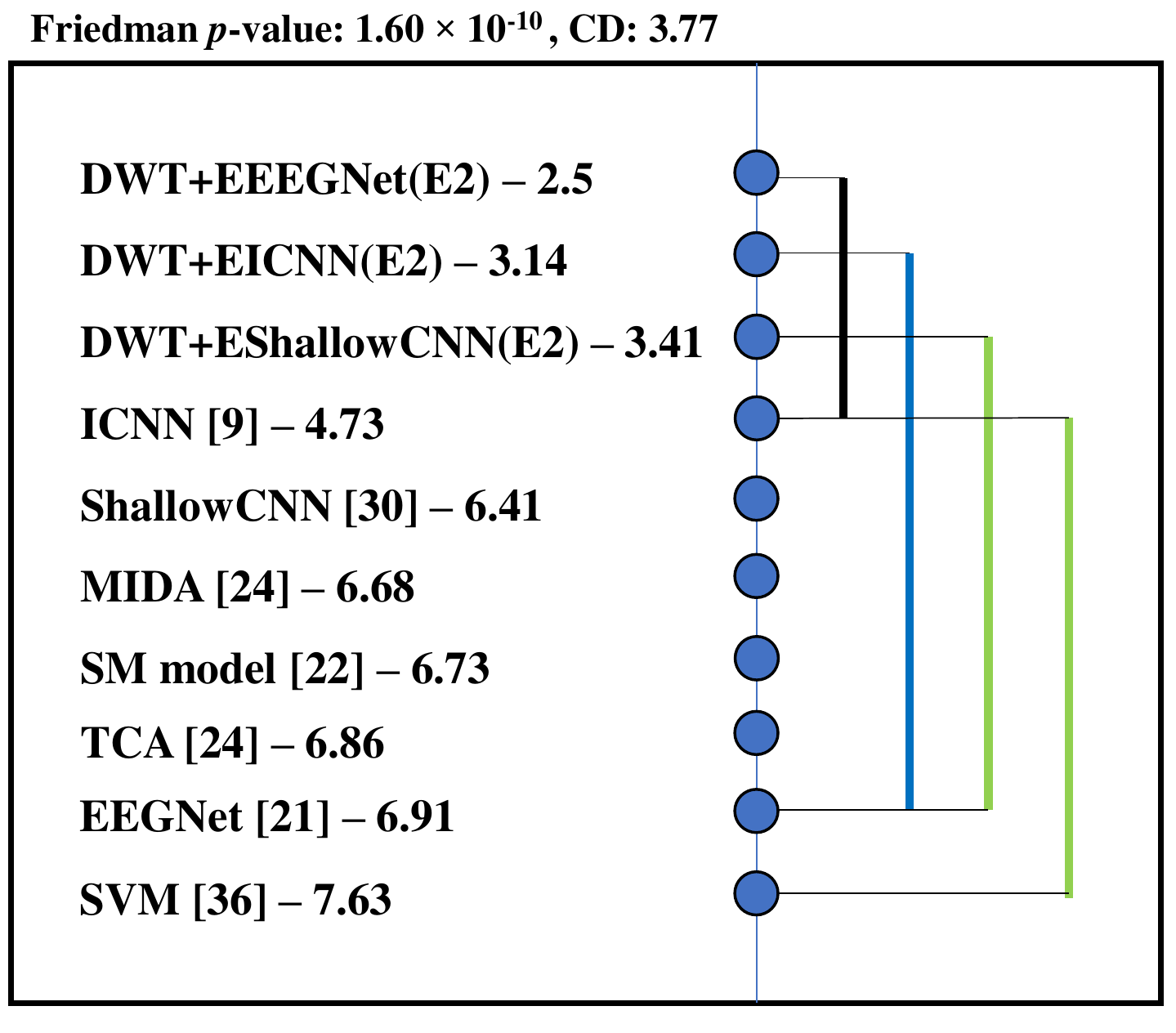}
  \caption{Fredman and Nemenyi test results}
  \label{CD}
\end{figure}

To rank the proposed ensemble models with different backbone architectures and the baseline methods, Friedman and Nemenyi \textit{post hoc} tests were performed in terms of LOSO average accuracy. For each backbone network discussed earlier (\textit{i.e.}, ICNN, EEGNet, and ShallowCNN), the ensemble models with the best performance were selected for ranking. Based on the calculated critical distance (CD) of 3.77, the results were shown in Figure \ref{CD}. All ensemble models of different architectures achieved a higher rank than the baseline methods. Overall, the results of LOSO average accuracy and ranking demonstrated that the proposed framework is suitable for any CNN architecture, with deep CNN having a slightly better performance than shallow CNN. Moreover, the Wilcoxon signed-rank test was conducted to further compare the performance of the DWT-based ensemble EEGNet in E2 mode (Ranked 1) and the DWT-based ensemble ICNN in E2 mode (Ranked 2). Results showed that the DWT-based ensemble EEGNet in E2 mode has a significantly better performance than the DWT-based ensemble ICNN in E2 mode ($p<0.05$). Hence, the DWT-based ensemble EEGNet in E2 mode achieved the SOTA performance. 

\subsection{Investigation on the effect of component-specific batch normalization in proposed framework}
\textcolor{black}{To illustrate the advantage of the introduced CSBN layer, the proposed models were compared with their variants in which the CSBN layer was replaced with the BN layer. The comparison results were presented in Table \ref{table:BN Comparison}. It could be observed that the CSBN layer substantially improved the performance of all proposed models compared with simply applying the BN layers. The main reason for this improvement is the ability of the CSBN layer to reduce subject variability.}

\textcolor{black}{Subject variability was regarded as a domain shift problem which means the distributions of different subjects may be different \cite{8852347}. In the BN layers, the statistics, mean and standard deviation, are usually calculated based on the training data's distribution. The obtained statistics are then directly applied to the testing data during testing. However, in the cross-subject EEG-based classification tasks, the statistics of the training data cannot precisely reflect that of the testing data. This may lead to the generation of inferior hidden features during testing and deteriorate the testing performance. Therefore, the CSBN layer is beneficial, of which the mean and the standard deviation of the hidden features are automatically calculated for the model of each component during testing instead of adopting from the training sets. This allows the feature alignment between the training data and the testing data in different complexity levels. As a result, the effectiveness of the CSBN layer designed for reducing subject variability and boosting the performance of the proposed ensemble framework was illustrated.}

\textcolor{black}{On top of that, based on Table \ref{table:BN Comparison}, it is worth emphasizing the effectiveness of decomposition and ensemble learning in the proposed framework on performance improvement again, as highlighted by the higher LOSO average accuracy of the ensemble models that used decomposed components with CSBN than that of ICNN that used the original EEG signals with AdaBN.}  

\begin{table}[]
\centering
\caption{LOSO average accuracy (\%) of using BN and CSBN in the proposed framework. EICNN: ensemble ICNN.}
\label{table:BN Comparison}
\begin{tabular}{cccc}
\hline
             & \textbf{\textit{w}} BN  & \textbf{\textit{w}} AdaBN & \textbf{\textit{w}} CSBN \\ \hline
ICNN \cite{9714736}        & 76.13 & 78.35   &        \\
VMD+EICNN+E1 & 76.11 &         & 78.50  \\
EWT+EICNN+E1 & 76.37 &         & 79.15  \\
EMD+EICNN+E1 & 77.04 &         & 80.27  \\
DWT+EICNN+E1 & 79.74 &         & 81.58  \\
VMD+EICNN+E2 & 78.02 &         & 78.97  \\
EWT+EICNN+E2 & 78.21 &         & 81.01  \\
EMD+EICNN+E2 & 78.05 &         & 81.22  \\
DWT+EICNN+E2 & 79.84 &         & 82.11  \\ \hline
\end{tabular}
\end{table}

\subsection{Analysis of ensemble learning in proposed framework}

\begin{figure}[htbp]
  \centering
  \subfigure[Results of EMD-based models]{
  \label{EMD}
  \includegraphics[width=0.45\textwidth]{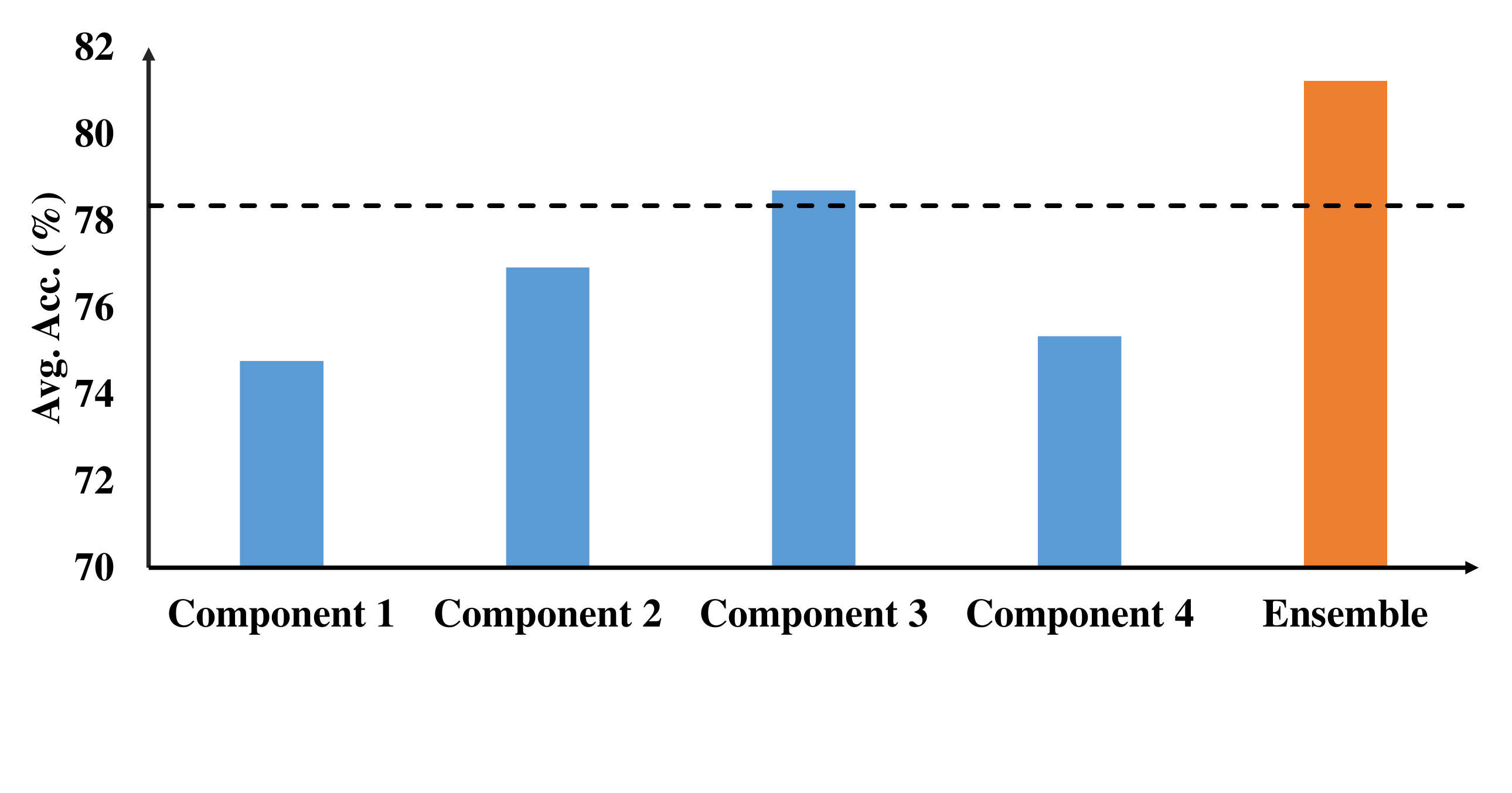}}
  \subfigure[Results of DWT-based models]{
  \label{dwt}
  \includegraphics[width=0.45\textwidth]{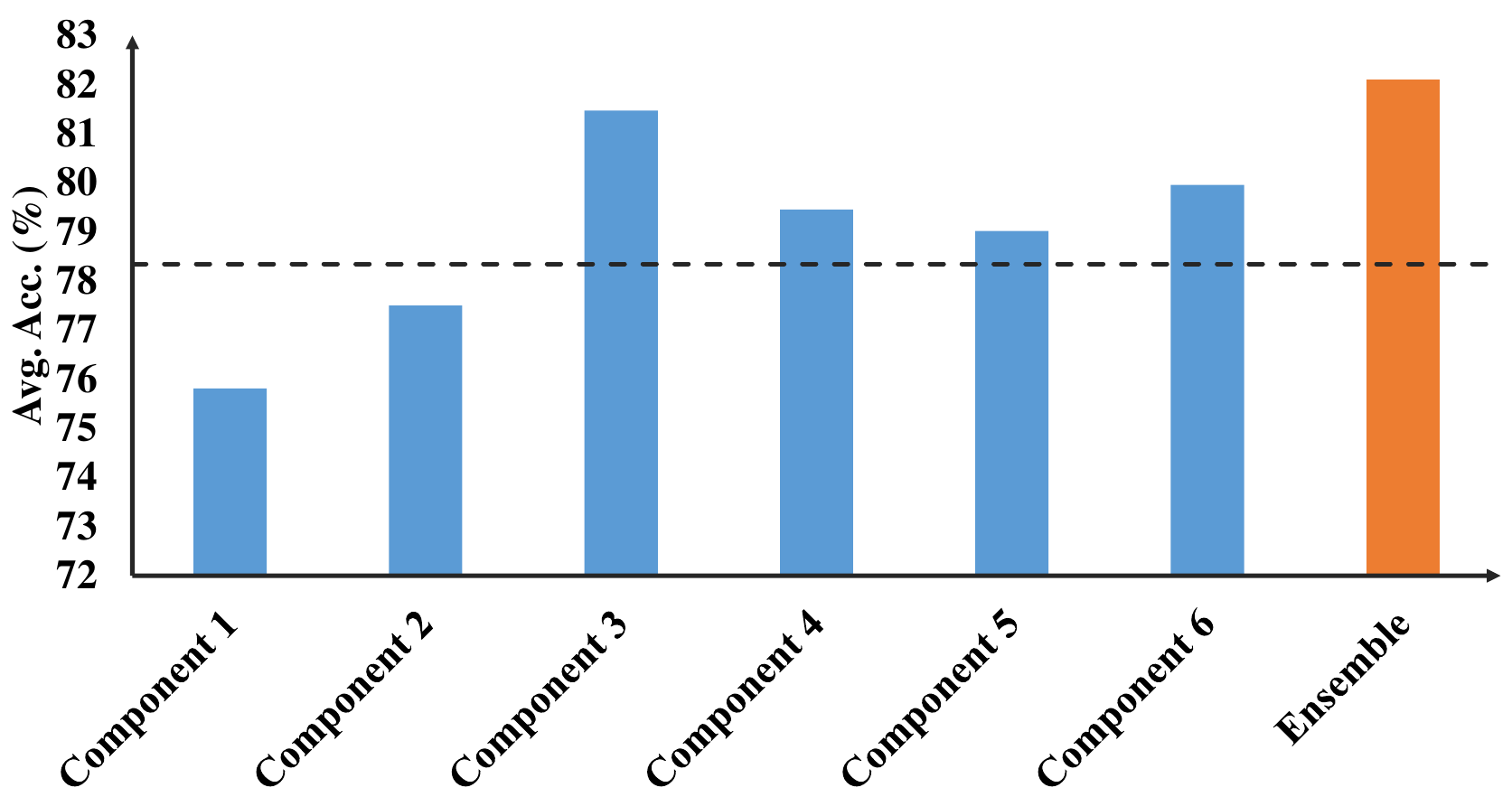}}
  \subfigure[Results of VMD-based models]{
  \label{vmd}
  \includegraphics[width=0.45\textwidth]{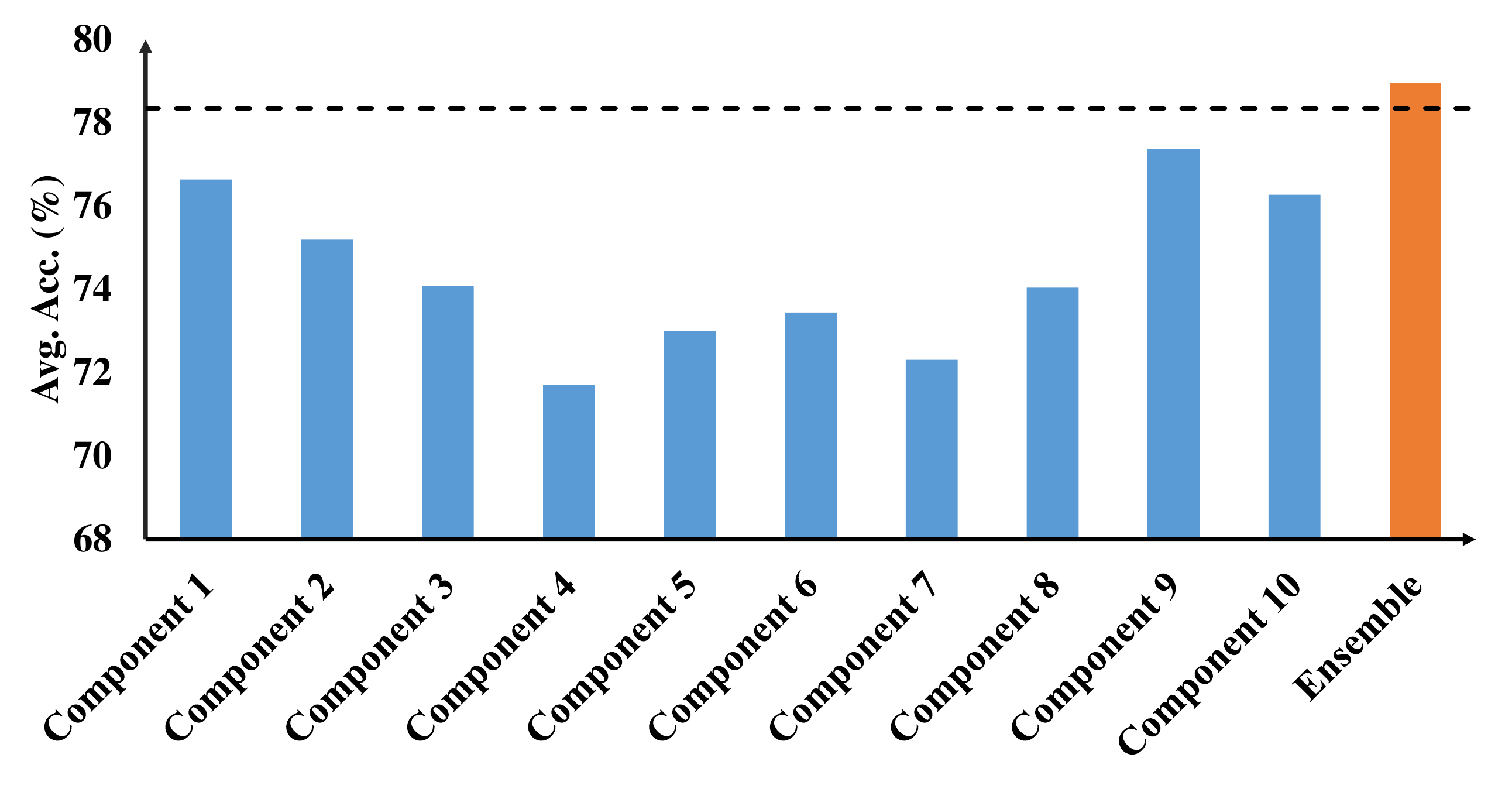}}
  \subfigure[Results of EWT-based models]{
  \label{EWT}
  \includegraphics[width=0.45\textwidth]{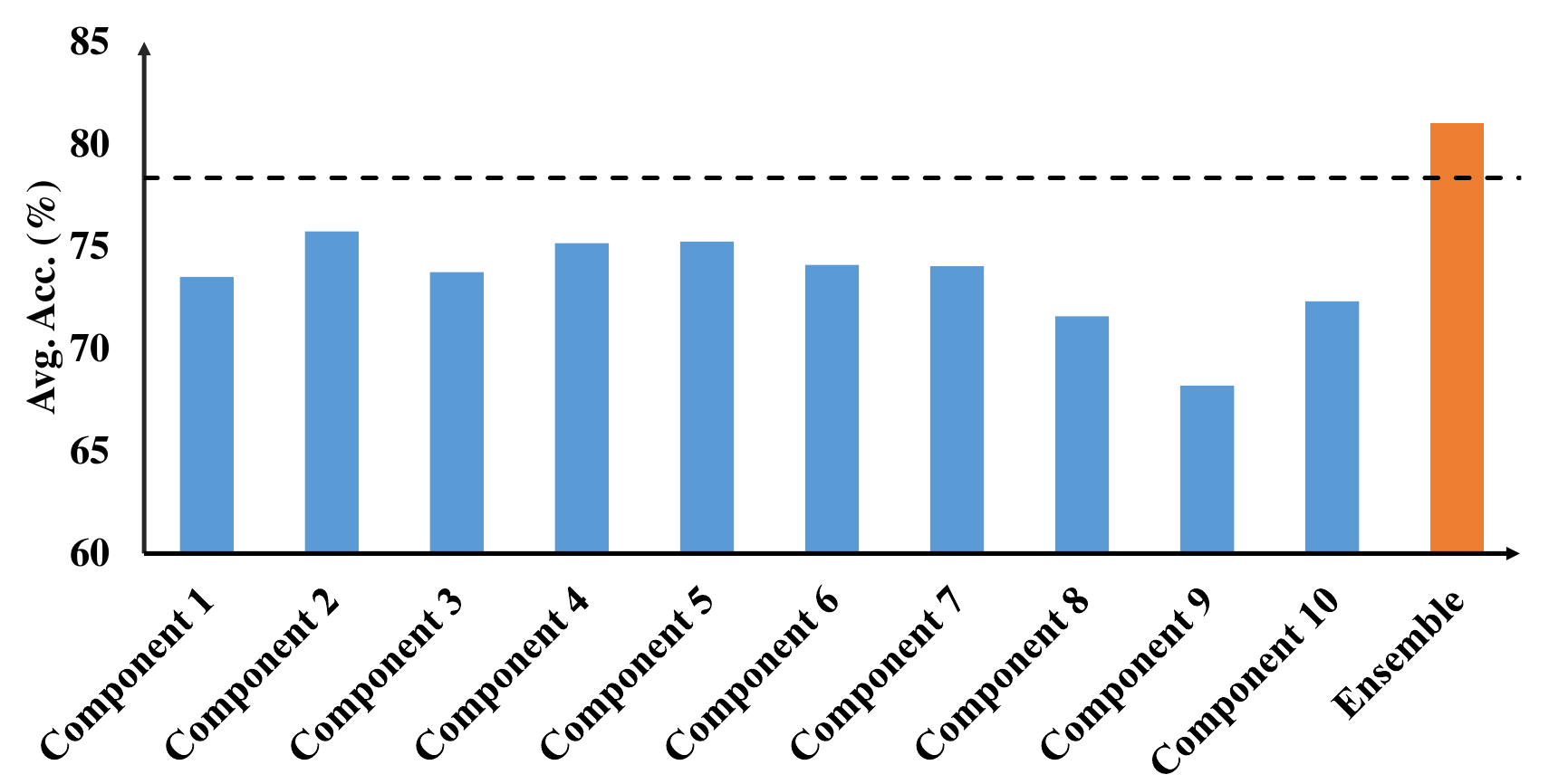}}
  \caption{LOSO average accuracy (\%) of ensemble learning and models trained individually on different decomposed components. The black dotted line represents the average accuracy of baseline, ICNN, trained on the original EEG signals. }
  \label{fig:each decomposed component}
\end{figure}

\textcolor{black}{The effectiveness of ensemble learning is investigated in this section. Specifically, we compared the performance of three groups, namely the model trained with the original EEG signals, the models trained with each decomposed component and the results of ensemble learning. It was shown in Figure \ref{fig:each decomposed component} that ensemble learning  could comprise the outputs of all models trained individually on the different decomposed components. For all the decomposition methods, ensemble learning obtained a better performance than the baseline, ICNN, which trained on the original EEG signals, as well as the models trained individually on the different decomposed components. It is worth noting that while some components of less useful frequency bands may substantially deteriorate the performance when individually trained, ensemble learning in the proposed framework can effectively reduce the potential impact of these less beneficial components on the final recognition performance when their outputs are combined with those of more beneficial components in ensemble learning.} 

\textcolor{black}{Overall, the elements in the proposed framework work synergistically to boost the performance of fatigue recognition tasks. As the entire complex EEG signals are decomposed into components of different frequency bands, it may facilitate the model to focus the learning on specific frequency bands and extract features that are more beneficial for driver fatigue recognition as there will be less impact from the less useful frequency bands. Subsequently, while  decomposition allows the model to capture more distinguishable patterns, ensemble learning ensures that the amount of information used for fatigue recognition tasks is not compromised.}

\subsection{Visualization}
\begin{figure}[htbp]
  \centering
  \subfigure[Alert recognition]{
  \label{dwt_alert}
  \includegraphics[width=0.9\textwidth]{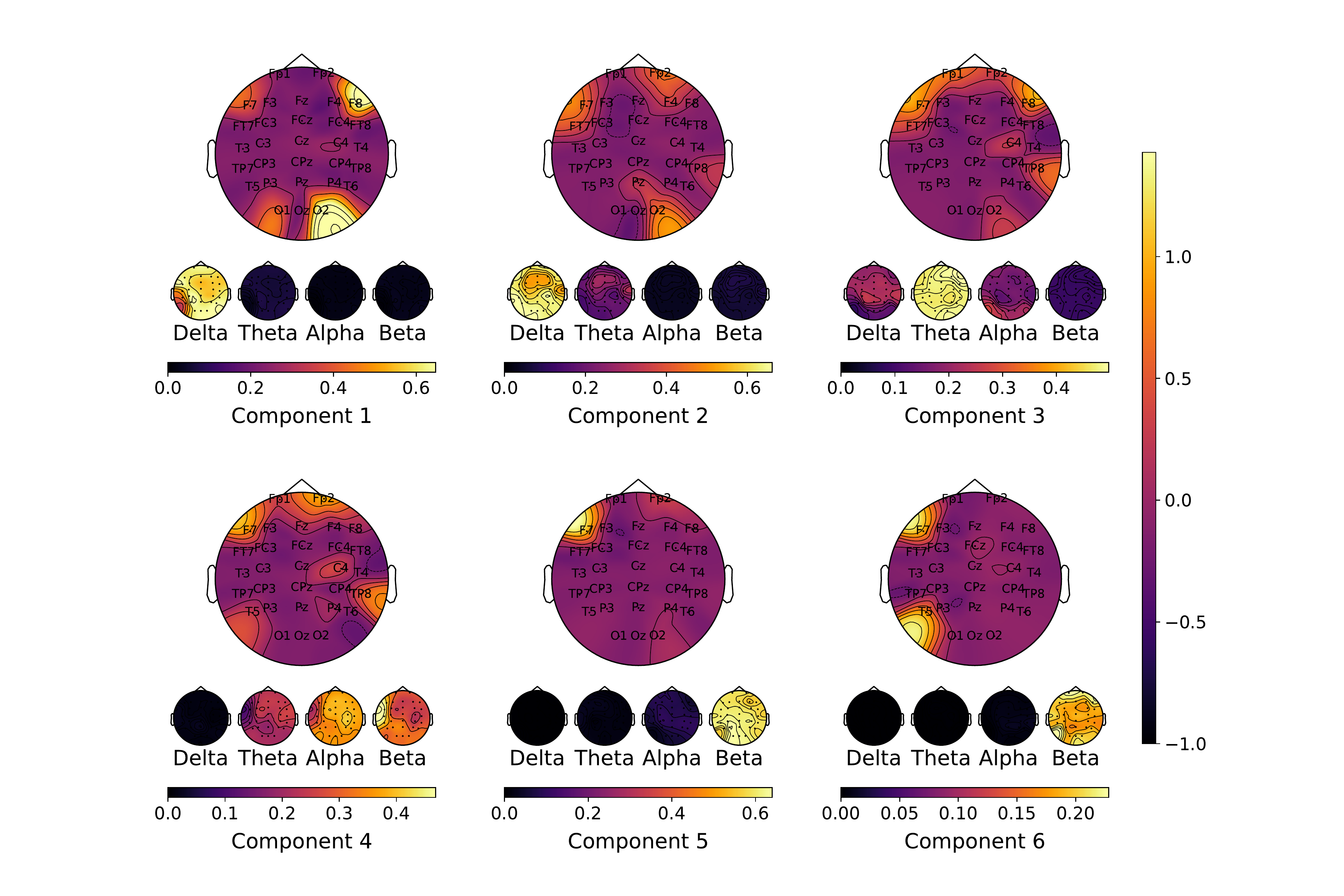}}
  \subfigure[Fatigue recognition]{
  \label{dwt_fatigue}
  \includegraphics[width=0.9\textwidth]{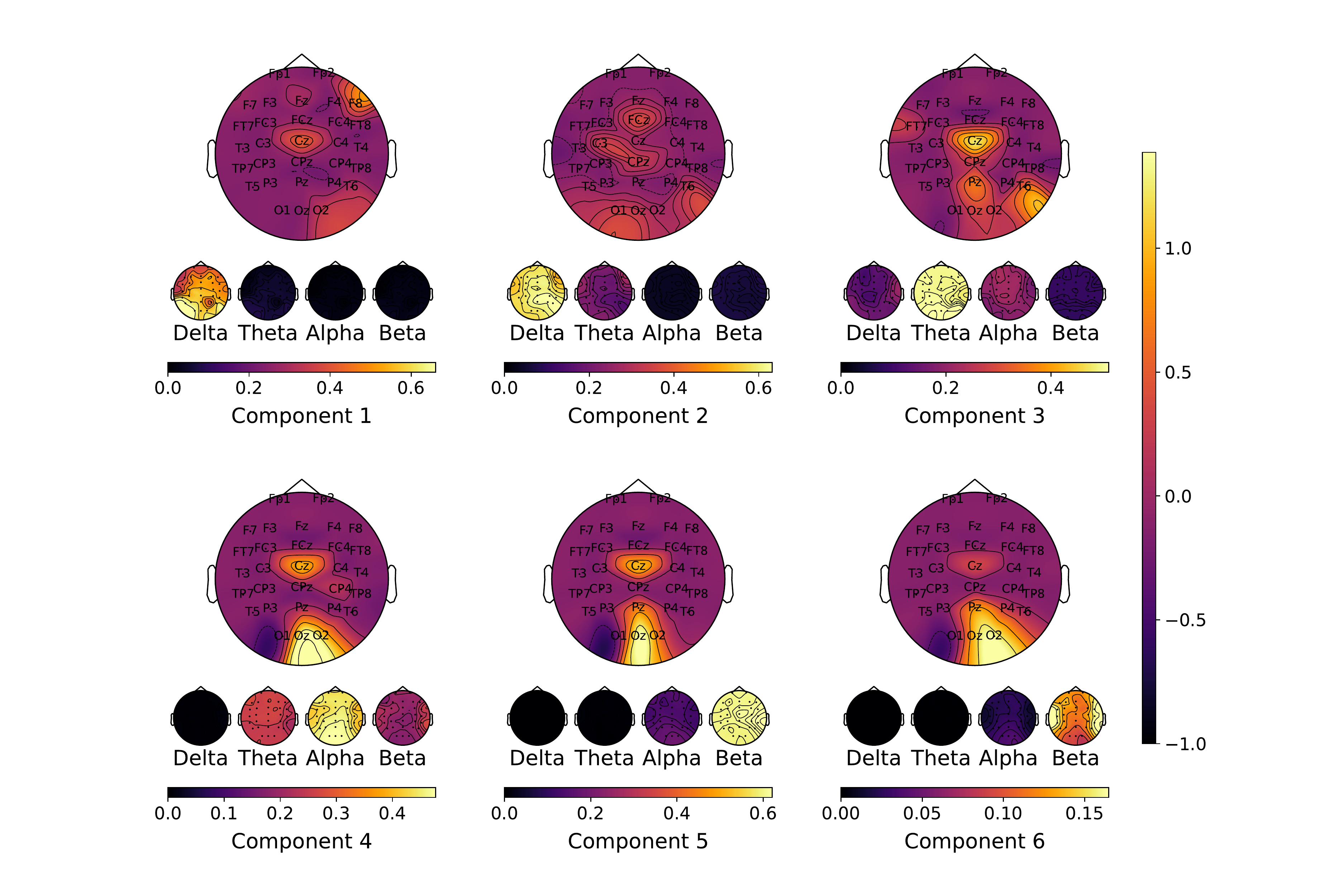}}
  \caption{Visualization of responsible regions and frequency bands based on DWT-based ICNN in E2 mode for (a) alert recognition and (b) fatigue recognition}
  \label{dwt_vis}
\end{figure}

\textcolor{black}{To better understand the responsible regions and frequency bands for classification in the proposed framework, a class activation mapping (CAM)-based model interpretation technique \cite{9714736} was employed to perform an interpretation for the used models. To use this technique to visualize the useful parts of all decomposed components simultaneously, we used the output of each individual model to calculate the weight of the contribution of each component to the final output. Since the DWT-based model in E2 mode showed the best performance among all decomposition-based models, we mainly presented the visualization results of the DWT-based ensemble ICNN in E2 mode. The visualization results were shown in Figure \ref{dwt_vis}. For each component, there were a topological heatmap of all channels above and the relative powers of the delta, theta, alpha, and beta frequency
bands for each EEG channel of the input component below. The topological heatmap summarized the degree that each channel contributes to the final
classification. The color bar on the right was for the topological heatmaps.}

\textcolor{black}{Based on the visualization results, it was observed that the occipital lobe and the frontal lobe with the beta frequency and the delta frequency band contributed more to alert recognition. Moreover, the regions of the centroparietal and occipital EEG channels with the theta, the alpha and the beta frequency bands mainly contributed to fatigue recognition. The finding on the responsible area was compatible with previous studies \cite{9714736}\cite{pal2008eeg}. Regarding the responsible frequency bands, the theta and alpha frequency bands have been found to be strong indicators of early fatigue and used in various driving simulator studies to identify the driver fatigue \cite{britton2016electroencephalography}\cite{SIMON20111168}. The delta frequency band has also been found to be the evidence of alert \cite{9714736}. Interestingly, the beta frequency band which is frequently associated with active concentration \cite{147683008X301478} was shown to be responsible for both alert and fatigue recognition. Overall, the visualization demonstrated that the proposed framework could focus on useful areas and useful frequency bands, illustrating its effectiveness.}

\section{Conclusion}
\textcolor{black}{Against the challenge of decoding non-stationary and high-complexity EEG signals for driver fatigue recognition, we propose a decomposition-based hybrid ensemble CNN framework that can capture more beneficial features from EEG signals while not compromising on the amount of information used for recognition tasks.} In the framework, a CSBN layer is added to reduce subject variability. The eight proposed models involving 4 decomposition methods and 2 ensemble output modes under our framework all showed superior LOSO average accuracy to the SOTA on the challenging cross-subject driver fatigue recognition task. Notably, DWT-based ensemble CNNs showed consistently better LOSO average accuracy than other decomposition methods-based ensemble models. Furthermore, two ablation studies were conducted to analyze the effect of total number of decomposed components for training and the use of different CNN architectures on the recognition performance of the proposed framework. Results demonstrated that in the range from 3 to 10, increasing the total number of decomposed components for training could generally boost the LOSO average accuracy of EMD-, EWT- and DWT-based models. Also, the framework could be extended to any architecture of CNN that have adequate decoding capability of EEG signals. In particular, deep CNN had better recognition accuracy and was found to be more suitable for decoding decomposed components in our framework. Lastly, ensemble learning which comprised the outputs of all individual models was effective in boosting the LOSO average accuracy by reducing the potential impact of the less beneficial components on the final recognition performance. \textcolor{black}{In conclusion, the proposed framework can significantly improve the EEG-based driver fatigue recognition performance. This can provide more directions for driver fatigue monitoring systems. In future works, more decomposition methods will be further investigated for driver fatigue recognition tasks.}

\bibliography{elsarticle-template}

\end{document}